\documentstyle[aps,floats,prl,twocolumn]{revtex}
\begin{document} 

\title{ Luther-Emery Stripes, RVB Spin Liquid Background and High $T_c$ Superconductivity
      }
\author{WenJun Zheng}
\address{ Department of Physics and Center for Material Research, Stanford University, Palo Alto, California 94305}
       
\date {Received 1 June 1999}       
\maketitle 

\begin{abstract}
         The stripe phase in high $T_c$ cuprates is modeled as a single stripe coupled to 
the RVB spin liquid background by the single particle hopping process. In normal state, the strong 
pairing correlation inherent in RVB state is thus transfered into the Luttinger stripe and 
drives it toward spin-gap formation described by Luther-Emery Model. The establishment of 
global phase coherence in superconducting state contributes to a more relevant coupling to
  Luther-Emery Stripe and leads to gap opening in both spin and charge sectors. Physical  
consequences of the present picture are discussed, and emphasis is put on the unification of 
 different energy scales relevant to cuprates, and  good agreement  is found with 
the available experimental results, especially in ARPES.

\end{abstract} 

   The universal presence of phase separation in high $T_c$ cuprates has been 
confirmed by  extensive  experiments, including elastic and inelastic neutron scatterings\cite{NS},  NMR and NQR\cite{NMRNQR}, and Angular Resolved Photon Emission Spectroscopy \cite{ARPES1}(ARPES) in $La_{2-x}Sr_xCuO_4$, $YBa_2Cu_3O_{7-x}$ and
 $Bi_2Sr_2CaCu_2O_{8+\delta}$ (Bi-2212) etc. The emerging picture is , upon hole doping beyond  $x=0.06$, quarter-filled \cite{footnote1} hole rich stripes begin to form, separating the 
 copper oxide plane into slices of antiferromagnetic insulating regions,
 with the inter-stripe distance in proportion to $1/x$, where $x$ is the density of 
doped holes. Above $x=1/8$ and  inside the overdoped regime, incommensurate stripe modulation persists, although the inter-stripe spacing saturates, with 
the excessive holes overflowing into insulating regions , signifying  the  crossover to
conventional metallic phase with overall homogeneity. Besides, the stripes are dynamically fluctuating and may coexist with superconductivity. Dated back to the late 1980's, the relevance of phase separation and dormain walls
 to high $T_c$ cuprates was already under considerable discussions\cite{EarlyStripe}. In 1993, Emery and Kivelson 
 suggested a scenario  of mesoscopic phase separation frustrated by long-range Coulomb interactions \cite{KE1}, as a general consequence of doping a strongly correlated insulator, and   they also pointed out the relevance of dynamical stripes to high $T_c$ superconductivity \cite{KE2}. The origin of phase separation is still under hot debate. Another equally important issue that will be treated here is: assuming the presence of stripes coupled to an undoped background, can we improve our understanding of  the interesting and even puzzling physical features revealed in
 both normal state and superconducting state of cuprates? The importance of such exploration has been recently emphasized in \cite{Neto1}\cite{Others3}, and some 
interesting results have been reported . Most of these attempts treat the stripe as
 a 1D or quasi-1D Luttinger Liquid coupled to neighbouring stripes \cite{Others1} or insulating background which is either modeled as a canonical antiferromagnet \cite{Others2} or as another 1D Luttinger Liquid  \cite{Others3}. The couplings through  pair tunneling \cite{Others3} or spin exchange \cite{Others2} have been discussed. Here, in contrast to \cite{Others3} and \cite{Others2},
 I emphasize the importance of coupling a stripe to a truly anomalous 2D insulating background, which  has its hidden unconventional nature of RVB (Resonating Valence Bond ) spin liquid, under the  classical apparel of  antiferromagnetic (AFM) order. It is shown that , through single particle hopping \cite{footnote3} between
 1D stripes and 2D RVB background, a normal state pseudo-gap $\Delta_n$ is "induced" inside the stripe's spin sector, which coincides with the mechanism of spin gap formation in a class of 1D electron systems named after Luther and Emery \cite{LEModel} . Further more, inside the superconducting state, the presence of global phase coherence in RVB order parameter contributes to an even more relevant pairing coupling to 1D stripe, which results in a gap of   $\Delta_{sc}$ opening in both spin and charge sectors.  Experimental consequences of 2 quantitatively different gaps are discussed, and good agreement is found with ARPES results \cite{White}. 

      The brilliant idea of RVB states was advanced by Anderson soon
 after the discovery of high $T_c$ superconductivity \cite{RVB}. The RVB state is described by a coherent superstition of different configurations of valence bonds, which was expected to 
 be a reasonable approximation to the ground state of insulating spin 1/2 Heisenberg Model, especially with frustration or hole doping, although the ground state of undoped cuprates clearly  has a Neel order.  Lately there has been renewed interest in the plausible relevance of RVB correlation to cuprate physics at relatively high energy scale, motivated both experimentally and theoretically. Recent ARPES result on $Ca_2CuO_2Cl_2 $ by F. Ronning et al. \cite{Ronning} reveals the presence of a d-wave dispersion along the remnant Fermi surface and a Dirac like dispersion isotropically focused around $(\pi/2, \pi/2)$, which is exactly what was predicted for the "$\pi$-flux phase" \cite{Piflux} of RVB spin liquid, where  $\epsilon(k)=J \sqrt {\cos ^{2} k_x +\cos ^{2} k_y}$, and can not be described within the spin density wave picture although the latter can account for the low-lying spin excitations in the Neel state.  Further numerical results also support the presence of a local RVB spin liquid state around a doped hole with momentum $\bf{k}=(\pi,0)$\cite{SpinLq}, accompanied by an anti-phase of spins around the hole which may be relevant to the generation of anti-phase domain walls in striped phase of cuprates. Theoretically, Kim and Lee show that Neel order can be restored in $\pi$-flux phase through dynamical mass generation of gauge fluctuations at low temperature \cite{Massgen} , which points  toward an  emerging consistent RVB picture spanning from ground state to high energy scale physics \cite{footnote2}.   
 Based on the above results, I suggest that one can model the environment of a quarter-filled stripe as a RVB spin liquid, which is coupled with the stripe  by a single particle hopping term that conserves the momentum along the stripe direction . To get started, one can first ignore the inter-stripe correlation in the normal state of underdoped cuprates, considering the strong incoherence revealed by experiments.
    The total Hamiltonian is given by 
\begin{eqnarray}
              H(c,c^+,d,d^+)&=&H_{1D}(d,d^+) + H_{RVB}(c, c^+)   \\   \nonumber
   ~~~~~~~~~~~~                    &+&H_{couple}(c,c^+,d,d^+) ,
\end{eqnarray}
 where $c$, $c^+$ and $d$, $d^+$ represent the annihilation and creation operators of a single particle in 2D RVB background and 1D stripe, respectively. 
$ H_{couple}(c,c^+,d,d^+)=\sum _{{\bf{k}},q,\sigma}V c^+_{\bf{k},\sigma} d_{q,\sigma} \delta_{k_x,q} +h.c.  $,
where only horizontal stripe is considered, ${\bf{k}}=(k_x,k_y)$, and momentum conservation is ensured by requiring
$k_x=q$. $V$ gives the hopping matrix element, which is vital in deciding  different energy scales relevant to cuprates , as will be discussed later \cite{footnote3}.

A routine Hartree-Fork decoupling is applied to the  RVB Hamiltonian $H_{RVB}$ \cite{RVB},

\begin{eqnarray}
            H_{RVB}&=&-J\sum _{<ij>}b^+_{ij} b_{ij} \\ \nonumber
                   &=&-J\sum _{<ij>}(\Delta^*_{ij} b_{ij} +\Delta_{ij} b^+_{ij}-|\Delta_{ij}|^2),
\end{eqnarray}
where $b^+_{ij}=\frac{1}{\sqrt{2}}[c^+_{i,\uparrow}c^+_{j,\downarrow}-c^+_{i,\downarrow}c^+_{j,\uparrow}]$, $\Delta_{ij}$ is
 the RVB order parameter defined on each bond between 2 nearest neighbors, which is reduced to the mean field average of $b_{ij}$ operator at the saddle point level.
Then one can change into the momentum space,  that is
\begin{eqnarray}
      H_{RVB}&=&\frac{-J}{\sqrt{2V}} \sum _{\bf{k},\bf{q}}[\Delta^*_{\bf{k},\bf{q}}(c_{\bf{q},\downarrow}c_{\bf{k}-\bf{q},\uparrow}-c_{\bf{q},\uparrow}c_{\bf{k}-\bf{q},\downarrow})+h.c.],
\end{eqnarray}
where $\Delta_{\bf{k},\bf{q}}=\sum_{\bf{\hat{\delta}}} \Delta_{\bf{k},\bf{\hat{\delta}}} e^{i \bf{q}\bf{\hat{\delta}}}$,   and 
$\Delta_{\bf{k},\bf{\hat{\delta}}}=\frac{1}{\sqrt V} \sum_{\bf{r_i}} \Delta_{i,i+\bf{\hat{\delta}}} e^{-i \bf{k} \bf{r_i}}  $  ($\bf{\widehat{\delta}}=\pm  \widehat{x},\pm \widehat{y}$).

There are many possible mean field states in RVB theory \cite{Fradkin}, among which the "$\pi$-flux
 phase" is selected here, because of its low energy, conservation of time-reversal symmetry and possible connection to AFM long range order\cite{Massgen}.
 In "$\pi$-flux phase", $\Delta_{ij}=\Delta_0 e^{i\phi_{ij}}$ is chosen to have uniform amplitude $\Delta_0$, while its phase $\phi_{ij}$ is selected to ensure that staggered  $+ \pi$ and $-\pi$ flux is threaded through each plaquette. For convenience, one can choose $ \phi_{ij}= \pm \pi/4$.
   Therefore, $H_{RVB}$ is simplified to

\begin{eqnarray}
      H_{RVB}&=&-J\sum _{\bf{q}}[\Delta^*_0 \gamma({\bf{q}})(c_{{\bf{q}},\downarrow} c_{-{\bf{q}},\uparrow}-c_{{\bf{q}},\uparrow}c_{-{\bf{q}},\downarrow})+h.c.] \\   \nonumber
                   &+&i J\sum _{\bf{q}}[\Delta^*_0 \eta({\bf{q}})(c_{{\bf{q}},\downarrow} c_{{\hat{\pi}-\bf{q}},\uparrow}-c_{{\bf{q}},\uparrow}c_{{\hat{\pi}-\bf{q}},\downarrow})-h.c.],
\end{eqnarray}
where $\gamma({\bf{q}})=\cos {q_x} +\cos {q_y} $,  $\eta({\bf{q}})=\cos {q_x} -\cos {q_y} $ and
${\widehat{\pi}}=(\pi,\pi)$. 
Then perform Euclidean path integral over the 2D degrees of freedom and obtain the low energy effective  action for 1D stripe  as follows
\begin{eqnarray}
   e^{-S_{eff}}&=&\exp \{-\int_{0}^{\beta} H_{eff}(d,d^+)d\tau +\frac{1}{\beta} \sum_{n} i\omega_n d^+ d \}  \\   \nonumber
                      &=&\int_{}^{} d c d c^+ \exp\{-\int_{0}^{\beta} [H_{1D}(d,d^+) + H_{couple}(c,c^+,d,d^+)   \\ \nonumber 
                      &+& H_{RVB}(c,c^+)] d\tau + \frac{1}{\beta} \sum_{n} i\omega_n (c^+ c+d^+ d) \},
\end{eqnarray}
where $\omega_n=\frac{\pi n}{\beta}$ ($n$ is odd integer), $\beta=\frac{1}{k_B T}$. 

Assuming  $J|\Delta_0|>>V$ , one can change back to 1D coordinate system and get
\begin{eqnarray}
      H_{eff}&=&H_{1D}(d,d^+) -\frac{V^2}{16J }\sum _{l}[\frac{1}{\Delta^*_0}(d_{l,\downarrow} d_{l+1,\uparrow}-d_{l,\uparrow}d_{l+1,\downarrow}) \\   \nonumber
                  &+&h.c.] -i \frac{V^2}{16J}\sum _{l}[ \frac{(-1)^l}{\Delta^*_0} (d_{l,\downarrow} d_{l+1,\uparrow}-d_{l,\uparrow}d_{l+1,\downarrow})-h.c.] .
\end{eqnarray}
     
Then go to the continuum limit, $d_{l,\sigma}\rightarrow \sqrt{a}\Psi(x=la)_{\sigma}$,  with the size of unit cell $a~ \rightarrow 0$
and retain only the slow varying part of $H_{eff}$,  we finally arrive at the following correction to $H_{1D}$ due to its coupling to RVB background,
\begin{eqnarray}
\label{EQ1}
 \Delta H_{eff}=g \int_{}^{} dx \Psi_\downarrow(x)\Psi_\uparrow(x) + h.c.  ,
\end{eqnarray}
where $g=-\frac{V^2\cos{k_F}}{8J\Delta^*_0}$, and $k_F=\pi/4$ for quarter-filled stripe.
We note that the 1D anomalous propagator is induced in stripes through hopping $V$ by the strong  pairing correlation inherent to the RVB background. This mechanism is central to the pairing process among mobile carriers inside stripes, via which superconductivity becomes viable.  

Based on the above result, I will discuss both normal state and superconducting state ,
 respectively. Let us first come to the issue of normal state pseudo-gap , which is 
deemed as very important but remains controversial. Theorists are sharply divided in 
whether  it is precursor pairing or otherwise has nothing to do with pairing, but caused by proximity to quantum critical point of , for example AFM phase transition. To treat normal state here, one can take the strong phase fluctuation in RVB order
 parameter into account , while its amplitude is basically non-zero and much less fluctuating. 
     Therefore, one can integrate out the  phase of $\Delta_0$,  and 
get
$$H_{eff}=H_{1D}(d,d^+) +g_1\int_{}^{} dx \Psi^+_\uparrow\Psi_\uparrow \Psi^+_\downarrow\Psi_\downarrow , $$
where $g_1 \approx \frac{-g^2 a^2}{2v}$, and $v$ is the bare Fermi velocity, 
which can be treated with the standard bosonization technique \cite{Bosoniztion}as follows

\begin{eqnarray}
      H_{eff}&=&  \int_{}^{} dx \{ [(\frac{K_cu_c}{2}-\frac{g_1}{2\pi})\Pi_c^2 + \frac{u_c}{2 K_c}{(\partial_x \Phi_c)}^2 ] \\   \nonumber
                  &+&  [(\frac{u_s}{2}+\frac{g_1}{2\pi})\Pi_s^2 + (\frac{u_s}{2 }+\frac{g_1}{2\pi}){(\partial_x \Phi_s)}^2 ]    \\  \nonumber
                  &+& g_1 \cos (\sqrt{8\pi}\Phi_s)  \},
\end{eqnarray}
where $\Phi_c,~\Pi_c$,  and $ \Phi_s,~\Pi_s$  , are conjugated boson operators representing 
density fluctuations in charge and spin sectors of 1D Luttinger Liquid \cite{Bosoniztion}, respectively. $u_c$, $u_s$ are the corresponding propagating velocities, and $K_c$ is 
a parameter of interaction.

In terms of renormalization group formulation,  $g_1 \cos (\sqrt{8\pi}\Phi_s)$ in $H_{eff}$ is marginally relevant, which results in the opening of a spectral gap in spin sector
$$ \Delta_s \propto   \sqrt{|g_1|} \exp{(\frac{v}{2\pi g_1})},  $$ where $v$ is the bare Fermi 
velocity. $\Delta_s$ is here identified as 
 the normal state pseudo-gap $\Delta_n$ that leads to spectral weight depletion in low energy spin fluctuations and single particle spectrum,  while the charge excitations remain gapless, which give rise to metallic transporting  along the stripe. This is of the same principle as the early results by Luther and Emery in exploration of spin gap formation as an instability of Luttinger Liquid \cite{LEModel}. Further more, the effect of $g_1$ on charge sector is also physically important, it leads to $K_c>1$ \cite{footnote6}, which ensures that singlet superconducting  fluctuation dominates over CDW (charge density wave) correlation, and drives the system close to the opening of charge gap and superconducting phase transition that accompanies it ( as will be clarified later).

Now let's turn to the superconducting state. It is generally agreed that global phase coherence 
is established at $T<T_c$, so that strong phase fluctuation in $\Delta_{ij}=\Delta_0 e^{i\phi_{ij}}$ is quenched, and the relevant correction to $H_{1D}$ becomes Eq[\ref{EQ1}] itself with $\Delta_0$ replaced by its average magnitude. 

Then standard bosonization gives
$$\Delta H_{eff}=2 g d_u \int_{}^{} \cos(\sqrt{2\pi}\Theta_c) \sin(\sqrt{2\pi}\Phi_s) dx , $$

The scaling dimension of $\Delta H_{eff}$ is $\frac{1}{2}+\frac{1}{2K_c}  $, so it is generally
 relevant except for very strong repulsive interactions(i.e. $K_c<1/3$). Unlike the normal state case discussed before, in $\Delta H_{eff}$ both spin sector and charge sector are 
coupled together by a relevant effective interaction and spin-charge separation typical of 
a Luttinger Liquid is thus broken and this kind of " spin-charge recombination " may be relevant to 
the generation of well-defined quasi-particles in superconducting state\cite{Long}. Under scaling to lower energy, $2 g d_u$
 is renormalized to divergence, so $\Theta_c$ and $\Phi_s$ oscillate around stable equilibrium positions and gaps open in both spin and charge excitations, which leads to non-magnetic ground state dominated by singlet superconducting fluctuations. For clarity, 
let's discuss the special case of $K_c=1$ and $u_s=u_c$. Then $H_{1D}$ can be decoupled
 into 2 independent Sine-Gordon models of $\Phi_{\pm}=\frac{1}{\sqrt2}(\Theta_c \pm \Phi_s)$,
 corresponding to 2 branches of free massive fermions. In this case, both spin gap and charge
 gap are equal, that is $$ \Delta_c=\Delta_s\propto 2\pi |g| d_u\propto 
\frac{V^2}{J \Delta_0} . $$ In general, the effect of  small $|K_C-1| > 0$ is only to mix the above two branches together, while the qualitative picture of gap formation remains robust. Further more,
 at leading order , it is expected that $\Delta_{s,c} \propto\frac{V^2}{J \Delta_0} \propto |g_1|^{1/2}$ is a fairly good approximation to start with \cite{Long},
. One can associate this gap with the superconducting gap $\Delta_{sc}$,
 identified as the quasiparticle gap  measured for example by ARPES in superconducting state.

Provided with two quantitatively different energy scales $\Delta_n$ and $\Delta_{sc}$  derived above , one can 
explore their experimental consequences. It is emphasized that,  without considering the
 inter-stripe coherent couplings, $V$ represents the strength of local hopping
 between a single stripe and its insulating background (its range is limited by inter-stripe distance), through which the strong pairing interaction intrinsic
 to RVB spin liquid is "transfered" into  the stripe, and leads to gap openings in both normal state and superconducting state.  In going toward overdoped region,  because RVB correlation is 
significantly suppressed, the relevant energy scale $g$ is  reduced to $J\Delta_0$, instead of 
$V^2/J\Delta_0$ \cite{Long}. Because $\frac{\Delta_n}{\Delta_{sc}}\propto \exp{(\frac{-v}{\pi a g^2})}$, $\Delta_n$ is much more suppressed compared with $\Delta_{sc}$\cite{Long}, which is well consistent with the ARPES results \cite{White}, and extensive experimental evidences supporting the "absence" of normal state gap in overdoped region \cite{footnote5}. Besides, by combining the 
present scenario with the spectral properties of Luther-Emery system \cite{LESpectrum}
, one can understand the broad "edge" feature near ($\pi$,0) in ARPES of underdoped normal state, as due to the  proximity toward charge gap formation  that turns the power law singularity ($\propto \omega^{\alpha-1/2}$,$0<\alpha<<1/2$ ) into a non-singular edge in $A(k,\omega)\propto \omega^{\alpha-1/2}$ ($\alpha>1/2$) \cite{Long}. However, this singularity is restored in overdoped region where the effect of RVB background on stripe is 
much weakened, therefore singular peaks with long 
tails are preserved in  $A(k,\omega)$ spectrum, as is consistent with what was observed in  ARPES \cite{ARPES2} .

    In superconducting state, global phase coherence allows single particle hopping
 between adjacent stripes through higher order process. From the calculation of corresponding matrix element $t'\approx \frac{\hbar^2}{2m^* d^2}$($ d$ is inter-stripe distance) , one can extract  the effective mass $1/m^*\propto \frac{V^2}{J\Delta_0}$ (underdoped case), and thus estimate the Josephson coupling energy 
$E_J\approx \frac{\hbar^2 \rho_s}{2m^*d} \propto  \frac{\Delta_{sc}}{d},$
where $\rho_s$ is the superfluid density of a single stripe \cite{Long} . In underdoped region, one can 
attribute superconducting transition to the global phase ordering \cite{EK} and therefore $T_c \propto E_J\propto  x \Delta_{sc} $, which agrees well with two facts: first, $T_c \propto  x$;
second, $T_{c,max}$ scales with $\Delta_{sc}$ among the cuprates family.
 In overdoped region, $T_c\propto \Delta_{sc} \propto J\Delta_0$ because  $d$ is saturated and a new energy scale $J\Delta_0$ takes the place of $\frac{V^2}{J\Delta_0}$, this is consistent with the BCS like relation observed in overdoped cuprates .

Before end, three comments are in order. First, the present scenario opens new route toward the understanding of the subtle relation between pseudo-gap  and superconducting gap,  in that both $\Delta_n$ and $\Delta_{sc}$ have the same origin :
 strong pairing interaction in RVB background, but can be quantitatively different in their
 dependences on  $V$ and $J \Delta_0$. Secondly,  one can unify  the important energy scales : $\Delta_n$, $\Delta_{sc}$, $E_J$, $T_c$, by determining 
their unique dependences on a single parameter ($V^2/J\Delta_0$ in underdoped region and $J \Delta_0$ in overdoped region), this explains the material-independent scaling in $\Delta_{sc} : \Delta_n: T_{c,max}$ among cuprates family, while a single material-independent $J$  can not.
Thirdly,  one can treat the "heavy mass" issue raised recently in \cite{ShenBalatsky} within the present picture: in underdoped cuprates, $\frac{k_B T_c}{x}=\hbar v^*\propto \Delta_{sc} 
\propto V^2/J\Delta_0$ and is roughly doping-independent. It can be connected to the flat dispersion perpendicular to horizontal stripes ( $\Gamma$ to (0,$\pi$) direction), as suggested in \cite{ShenBalatsky}, and can be attributed to slow hole motion transverse to stripes\cite{Long},
which limits the achievement of higher $T_c$.

    In conclusion,  I model the  stripe phase in high $T_c$ cuprates as a single stripe coupled to 
the RVB spin liquid background by the single particle hopping. In normal state, the strong 
pairing interaction inherent in RVB state is therefore transfered into the Luttinger stripe and 
drives it toward  Luther-Emery Stripe with spin-gap formation. The establishment of 
global coherence in superconducting state contributes to a more relevant coupling to
   the stripe and leads to gap opening in both spin and charge sectors. Physical  
consequences of the present picture are discussed,  and  good agreement  is found with 
the available experimental results in ARPES.

     I thank S. A. Kivelson, Z. X. Shen , S. Doniach, D. L. Feng and J. P. Hu for discussions and comments on this work. The support from Stanford Graduate Fellowship (SGF) is acknowledged.


\begin{references}
\bibitem{NS}  J. M. Tranquada et al., Nature (London) {\bf 375}, 561 (1995); K. Yamada et al., 
Phys. Rev. B{\bf 57}, 6165(1998); G. Aeppli et al., Science {\bf 178}, 1432 (1997).
\bibitem{NMRNQR} A. W. Hunt et al., Phys. Rev. Lett.{\bf 82}, 
4300 (1999); F. C. Chou et al., Phys. Rev. Lett.{\bf 71}, 2323 (1993).
\bibitem{ARPES1} Z. X. Shen et al., Science {\bf 280}, 259 (1998).
\bibitem{footnote1} The possible existence of roughly quarter-filled stripes at hole density well above 1/8 was revealed by recent ARPES results, D.L.Feng et al.( to be published).
\bibitem{EarlyStripe} J. Zaanen and O. Gunnarson, Phys. Rev. B{\bf 40}, 7391 (1989); D. Poilblanc and T. M. Rice, ibid.{\bf 39}, 9749 (1989); H. J. Schulz, Phys. Rev. Lett.{\bf 64}, 1445 (1990); M. Inui and P. B. Littlewood, Phys. Rev. B{\bf 44}, 4415 (1991). 
\bibitem{KE1} S. A. Kivelson and V. J. Emery, in Proceedings of " Strongly Correlated Electronic Materials: The Los Alamos Symposium 1993", ed. by K. S. Bedell, et al ( Addison Wesley, Redwood City, 1994) p.619; V. J. Emery and S. A. Kivelson, Physica C {\bf 209}, 594 (1993).
\bibitem{KE2} V. J. Emery and S. A. Kivelson, LANL preprint cond-mat/9809083 (To be published in the proceedings of Stripes98).
\bibitem{Neto1} A. H. Castro Neto et al., Phys. Rev. Lett.{\bf 79}, 4629 (1997). 
\bibitem{Others3} V. J. Emery , S. A. Kivelson, and O. Zachar, Phys. Rev. B{\bf 56}, 6120 (1997).
\bibitem{Others1} A. H. Castro Neto and F. Guinea, Phys. Rev. Lett.{\bf 80}, 4040 (1998); 
A. H. Castro Neto, Phys. Rev. Lett.{\bf 78}, 3931 (1997).
\bibitem{Others2} Yu. A. Krotov, D. H. Lee and A. V. Balatsky, Phys. Rev. B{\bf 56}, 8367 (1997); M. Granath and H. Johannesson, Phys. Rev. Lett.{\bf 83}, 199 (1999).
\bibitem{footnote3} This is a more general starting point, because the 
interesting couplings like spin exchange and pair tunneling can be treated by going to the second order process. 
\bibitem{LEModel} A. Luther and V. J. Emery, Phys. Rev. Lett.{\bf 33}, 589 (1974); P. A. Lee,
  Phys. Rev. Lett. {\bf 34}, 1247 (1975).
\bibitem{White}  P. J. White et al., Phys. Rev. B{\bf 54}, 15669 (1996).
\bibitem{RVB} P. W. Anderson, Science {\bf 64}, 188 (1986); P. W. Anderson, in Frontiers and Borderlines in Many Particle Physics, Ed. by R. Schrieffer and R. A. Broglia (North Holland, Amsterdam, 1989).
\bibitem{Ronning} F. Ronning et al., Science {\bf 282}, 2067 (1998).
\bibitem{SpinLq} T.Tohyama et al, LANL preprint cond-mat/9904231.
\bibitem{Massgen}  D. H. Kim and P. A. Lee, Ann. Phys. {\bf 272}, 130 (1999).
\bibitem{footnote2} However, the discussion here concerning gap formation does not rely on the existence of such consistent picture. The relevance of RVB picture to excitations at relatively high energy scale (around J) is what is needed to justify the whole discussion. 
\bibitem{footnote4} Here for simplicity, a constant $V$ is adopted, which is good for the semi-quantitative discussion here. However, this treatment only partly captures the stripe's dynamical fluctuations through charge exchange with the 2D background (energy dissipation is 
albeit largely reduced because of the gapful RVB background coupled to stripes). In order to include the collective motion of stripes, collective momentum should be explicitly contained in a more sophisticated expression of $V$, which will be discussed in future publications.  
\bibitem{Piflux} G. Kotliar, Phys. Rev. B{\bf 37}, 3664 (1988); I. Affleck and B. Marston, Phys. Rev. B{\bf 37}, 3774 (1988). 
\bibitem{Fradkin} See Chapter 6, "Field Theories of Condensed Matter Systems", by E. Fradkin (Addison-Wesley Publishing Company, 1991).
\bibitem{Bosoniztion} For a review, see " Bosoniztion and Strongly Correlated Systems", by 
A. O. Gogolin, A. A. Nersesyan, and A. M. Tsvelik (Cambridge University Press, 1998).
\bibitem{footnote6} For stripes dominated by strongly repulsive interaction, the bare $K_c$
 can be close to 1/2 (as obtained in 1D Hubbard Model). In order to boost it beyond 1, we find
 it is necessary for the intra-stripe dispersion to be as small as $J\approx 130 mev$. Under this 
 assumption, the present formulation still applies.
\bibitem{Long}  W.J.Zheng et al, in preparation.
\bibitem{footnote5} It is commented, however, that the experiments up to now on ARPES, tunnelings
 and optical measurements do not seem to clearly resolve two energy gaps above and below $T_c$,
 for an experimental review, see T. Timusk and B. W. Statt, Rep. Prog. Phys. {\bf 62}, 61 (1999). 
\bibitem{LESpectrum} J. Voit, LANL preprint cond-mat/9602087.
\bibitem{ARPES2} Z. X. Shen and R. Schrieffer, Phys. Rev. Lett.{\bf 78}, 1771 (1997).
\bibitem{EK}V. J. Emery , S. A. Kivelson, Nature {\bf 374}, 434 (1997); 
\bibitem{NdLSCO} X. J. Zhou et. al. (to be published).
\bibitem{Tallon} J. L. Tallon et. al., Phys. Rev. Lett.{\bf 79}, 5294 (1997).
\bibitem{ShenBalatsky} A. V. Balatsky and Z.-X. Shen, Science {\bf 284}, 1137 (1999).


\end{references}
 \end{document}